\documentclass{article}

\usepackage[final,nonatbib]{nips_2018}

\usepackage[utf8]{inputenc} 
\usepackage[T1]{fontenc}    
\usepackage{url}            
\usepackage{booktabs}       
\usepackage{amsfonts}       
\usepackage{nicefrac}       
\usepackage{microtype}      

\usepackage{graphicx}
\usepackage{amsmath}
\usepackage{csquotes}

\newcommand{\R}{\mathbb{R}}
\newcommand{\K}{\boldsymbol{\Sigma}}
\newcommand{\nsigma}{\sigma_{\epsilon}}
\newcommand{\vect}[1]{\textbf{#1}}
\newcommand{\obs}[1]{\hat{#1}}
\newcommand{\pred}[1]{{#1}_{\star}}

\graphicspath{{Poster/}}

\title{A Bayesian take on option pricing with Gaussian processes}

\author{
  Martin~Tegnér\thanks{Currently with Department of Mathematical Sciences, University of Copenhagen and Oxford-Man Institute, University of Oxford.} \, and Stephen~Roberts\\
  Department of Engineering \& Oxford-Man Institute\\
University of Oxford\\
  Oxford, OX13PJ \\
  \texttt{\{mt,sjrob\}@robots.ox.ac.uk} \\
}

\begin{document}

\maketitle

\begin{abstract}
Local volatility is a versatile option pricing model 
due to its state dependent diffusion coefficient.
Calibration is, however, non-trivial as it involves  both proposing a hypothesis model of the latent function and a method for fitting it to data. 
In this paper we present novel  Bayesian inference 
with Gaussian process priors. We obtain a rich representation of the local volatility function with a probabilistic notion of uncertainty attached to the calibrate. We propose an inference algorithm and apply our approach to S\&P 500 market data.

\end{abstract}

\section{Introduction}

The local volatility 
model is widely used in finance within option pricing. 
One reason  is its  capability of reproducing any given surface of observed  prices from European options. 
The flexibility is due to a latent  function---the local volatility function---which uniquely defines the model's pricing operator. The crux of the matter is that it can only be unambiguously retrieved in the limit of infinite data. Numerous calibration approaches have therefore been suggested for its (re)construction. 
Deterministic approaches prevail, typically  parametric methods that involve steps of interpolation and extrapolation.

We look at  calibration from a probabilistic angle with an  approach based on Gaussian processes. This  gives a way of encoding prior believes about local volatility with a nonparametric model which is  flexible, yet prone to overfitting. 
 In addition to providing  point-estimate(s), we  draw posterior inference about local volatility to better understand  uncertainty attached with the calibrates. 
Our approach also gives means for probabilist prediction of local volatility as well as for deriving quantities from the model 
 which takes the posterior information into account.

\subsection{The local volatility model: theory and practice }
In mathematical finance, option pricing is often based on It\^{o}-process models  of the financial market. We consider the local volatility model \cite{derman1994,dupire1994pricing} where the  risk-neutral dynamics of a stock price $S$ is 
\begin{equation}\label{eqLVSDE}
dS_t = r S_t dt + \sigma(t,S_t)S_tdW_t,\quad t\geq 0,
\end{equation}
(see \cite{oksendal2003stochastic} for an introduction to stochastic differential equations). The latent {local volatility function} $\sigma:\R_+^2 \rightarrow\R_+$ is the essential component of \eqref{eqLVSDE} and our main target for modelling. $S$ is driven by a Wiener process $W$ and the constant $r$ is the  interest rate of a bank account with value $dB_t = rB_tdt$. 
Together, $(S,B)$ constitute the underlying financial market on which options are derivative.  

A European call option ({call} for short) with strike price $K$ and maturity time $T$ gives its holder a payoff  $\max(S_T-K,0)$ at maturity, where the terminal stock price $S_T$ is unknown at $t<T$. Thus, one need a method for computing the time-$t$ fair price $C(t,S_t,T,K)$ of the option. No-arbitrage pricing theory (e.g., \cite{bjork2009arbitrage}) yields that
\begin{equation}\label{eqRNVF}
C(t,S_t,T,K) = E\left[\left. e^{-r(T-t)} \max(S_T-K,0) \right| S_t \right] 
\end{equation}
where $E[\cdot|S_t]$ denotes conditional expectation given the current stock price $S_t$. Dupire \cite{dupire1994pricing} shows how to represent \eqref{eqRNVF} as the solution to a partial differential equation in the variables $(T,K)$
\begin{equation}\label{eqDupire}
\frac{\partial C}{\partial T} + rK\frac{\partial C}{\partial K} - \frac{K^2\sigma^2(T,K)}{2}\frac{\partial^2C}{\partial K^2} = 0
\end{equation}
with boundary condition $C(T=0,K) = \max(S_t-K,0)$.

\begin{figure}
\makebox[\textwidth][c]{  
\includegraphics[scale=0.45,trim=50 50 50 50,clip]{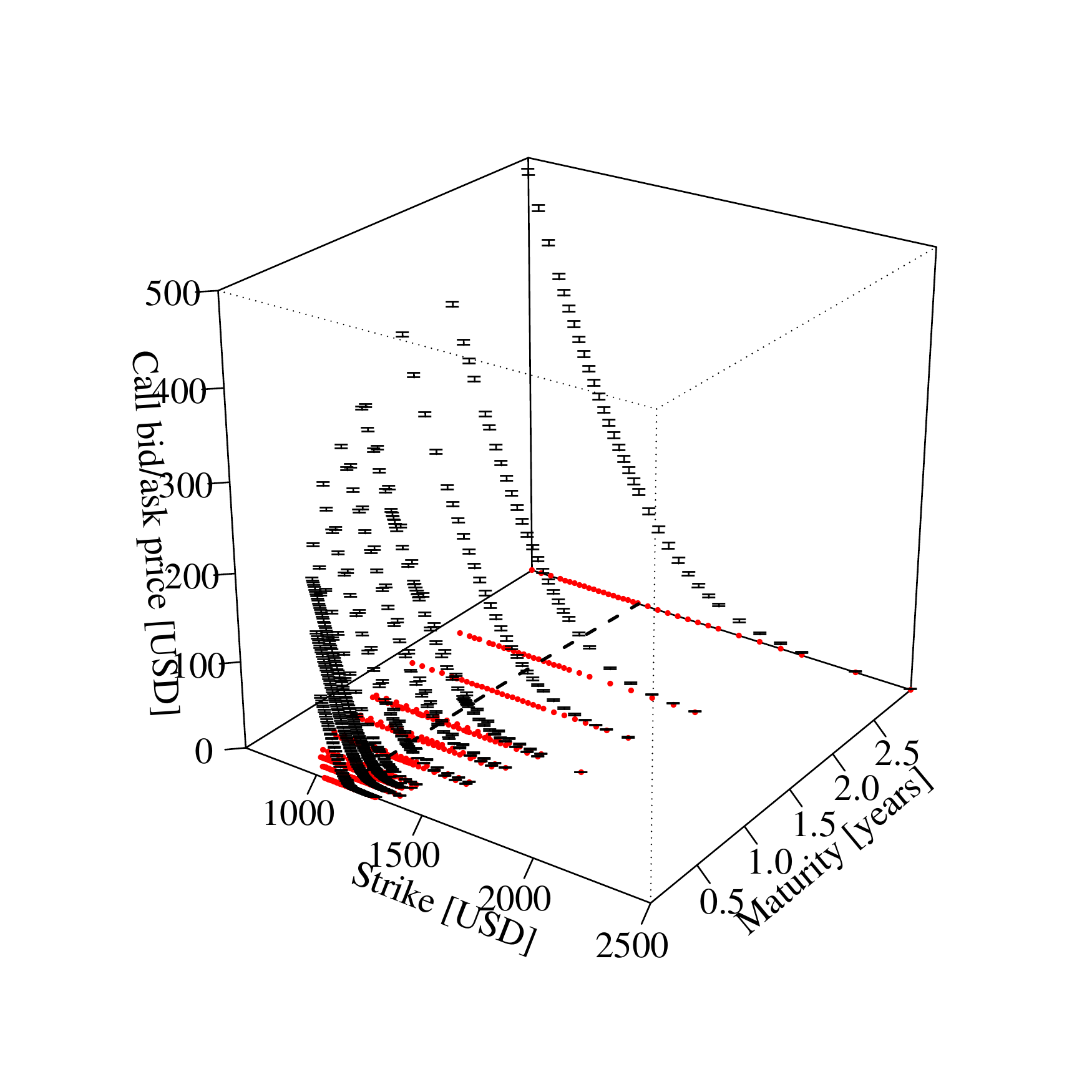}
\includegraphics[scale=0.45,trim=50 50 50 50,clip]{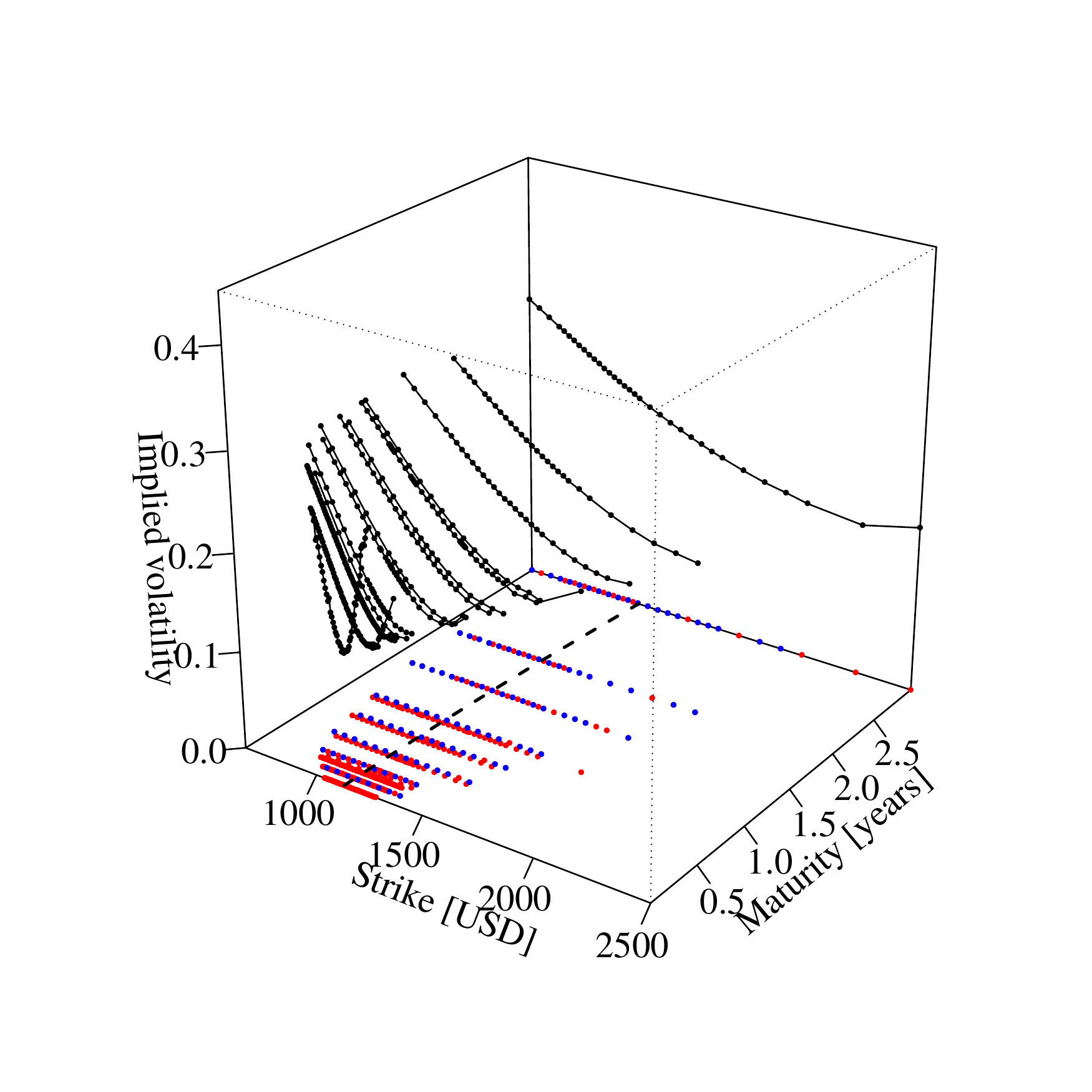}
}
\caption{\textbf{Left} Market prices of S\&P 500 call option as of 4 January 2010. The bid-ask price-interval for each quoted strike-maturity point (red dots sitting on the $x$-$y$ plane) is represented by a line-segment (\texttt{I}); 946 quotes in total. The dashed line shows the at-the-money strike level $K=S_t$. 
\textbf{Right} S\&P 500 implied volatility surface calculated from mid-market prices of the data in the left-hand plot. Blue dots indicate  strike-maturity points included in the  input set $\hat{\vect{x}}$ used for calibration. }
\label{fig1}
\end{figure}

In practice, an option-pricing model like \eqref{eqLVSDE} may be used for generating prices of  options not  quoted on the market,  deriving hedging strategies and  evaluating risk measures. Before employing a model for such purposes, it has to be fitted to data with a {calibration} approach  \cite{modelCalibration}. Here, data is market prices of liquid vanilla options--- European puts and calls---due to their role as benchmark instruments 
and since they are frequently (and publicly) traded. Figure \ref{fig1} shows the data we use for this paper.

A theoretical appeal with the local volatility model is that it can perfectly fit any given set of \textit{consistent} call (and/or put) prices. This means that there are no internal arbitrage opportunities in the set \cite{carr2005note}.  In practice, however, market prices are quoted only up to their bid-ask spread (left Figure \ref{fig1}) from which it is common to use   mid-prices for calibration. This is often formulated  as  a least-squares problem
\begin{equation}\label{eqG}
\inf_{\sigma} G(\sigma), \quad G(\sigma) =  \sum_{i=1}^{n} \left(C_i(\sigma) - \obs{c}_i	\right)^2
\end{equation}
where model prices $C_i(\sigma)$ correspond to mid-market prices $\obs{c}_i$.
When optimising \eqref{eqG} one needs to evaluate $C_i(\sigma)$, $\forall i$, for each candidate local volatility function. This can only be done with numerical methods. Dupire's equation speeds up this computation---thereby the overall calibration procedure---since a  numerical solution of \eqref{eqDupire} gives $C_{i}$  for all $i$ in a single sweep. Following \cite{gupta2014robust,cont2004recovering,jackson1998computation} among others, we  use a 
Crank–Nicolson scheme \cite{hirsa2012computational} for this purpose, i.e., for the mapping $\sigma\mapsto \{C_i(\sigma)\}_{i}$. 

\section{A probabilistic approach to nonparametric local volatility}\label{secPropLV}
We place  local volatility  in a Bayesian framework by a  probabilistic generative model of data. To this end, we take the space  of strikes and maturities,  $x:=(T,K)\in\mathcal{X} := \R^2_+$, to be the input space of a Gaussian process prior. The Gaussian process defines a distribution over functions from $\mathcal{X}$ to $\R$. 
It is uniquely specified by a covariance function $k_{\kappa}: \mathcal{X}\times\mathcal{X}\rightarrow\R$ parametrised by $\kappa$, and a mean function which we take to be zero \cite{rasmussen2006gaussian}. A  distribution over local volatility functions is  then defined through a positive transformation
\begin{equation}\label{eqGPprior}
 \sigma(x) = \exp f(x), \quad f(x) \sim \mathcal{GP}(0,k_{\kappa}(x,x')).
\end{equation}
The prior model \eqref{eqGPprior} provides means of encoding prior knowledge about local volatility. The stock price diffusion \eqref{eqLVSDE} and Dupire's equation \eqref{eqDupire} pertain such knowledge (to guarantee a solution of \eqref{eqLVSDE}  with continuous transition density) but it is only minimal: $\sigma:\mathcal{X}\rightarrow\R$ being continuous and positive suffices. Stronger regularity assumptions are often motivated from a practical viewpoint. A smooth volatility surface is desired when using the model for hedging and risk management, and to obtain a more robust calibration method. This is conveniently encoded by the squared exponential covariance 
\begin{equation}\label{eqSE}
k_{\kappa}(x,x') = \sigma_f^2\exp\left(-\frac{(T-T')^2}{2l^2_T}   \right)\exp\left(-\frac{(K-K')^2}{2l^2_K}   \right)
\end{equation}
since \eqref{eqGPprior}-\eqref{eqSE} has sample paths with continuous derivatives of all orders \cite{lindgren2012stationary}.

While the Gaussian process has the  title role in our  approach, we accommodate it with a hyperprior distribution over  $\kappa=(\sigma_f,l_T,l_K)$. These parameters specify the general shape of the functional output: $l_T,l_K$ control variability  along  input dimensions and  $\sigma_f$ controls output variance. 
Specifying sensible values for $\sigma_f,l_T,l_K$ is  less straightforward 
 and hyperparameter inference is therefore integral in the learning process. We assume independent sigmoid Gaussians 
\begin{equation}\label{eqSSG}
\kappa_i\sim p_h(\kappa_i) = \phi\left(\log(\kappa_i-1)-\log \kappa_{i,\max}\right)\frac{\kappa_{i,\max}}{\kappa_i(\kappa_{i,\max}-\kappa_i)},\quad \kappa_i\in (0,\kappa_{i,\max}),
\end{equation}
where $\phi$ is the standard normal density.  A limited support of the parameter space  make the model prone to identifiability issues and it  helps improving sampling efficiency \cite{aki}. We will also leverage that \eqref{eqSSG} defines a transformed Gaussian in the design of our sampling strategy.  

For the final component of the generative procedure we take the least-squares calibration \eqref{eqG} as basis for a likelihood. Market data $\mathcal{D}=\{\obs{c}_i,\obs{x}_i\}_{i=1}^{n}$ (the  \{mid-price, maturity, strike\} of quoted  options) is assumed to be \textit{fair} prices conditional on local volatility, and observed with additive Gaussian noise. In the simple case of independent noise with common standard deviation\footnote{We use boldface to denote a vector, or more generally, a set $\vect{x}=\{x_i\}_{i}$.} 
\begin{equation}\label{eqll}
\log p(\obs{\vect{c}}|\vect{f},\nsigma) = -\frac{1}{2\nsigma^2}\sum_{i=1}^n \left(  C(\vect{f},\obs{x}_i) - \obs{c}_i \right)^2 -\frac{n}{2}\log\left( 2\pi\nsigma^2\right).
\end{equation}
In the likelihood \eqref{eqll} we explicitly write out the dependency of the model price $C$ on a strike-maturity point $\obs{x}_i$ (corresponding to $\obs{c}_i$) and a finite set of $N\geq n$ volatility/functional values, $\vect{f}=\log\boldsymbol{\sigma}$, taken at an input set $\vect{x}$. For the finite difference scheme, the input set is a Cartesian product $\vect{x} = \{(T_i,K_j)\}_{i,j}$ and it is constructed to include all market inputs of $\mathcal{D}$, i.e., $\obs{\vect{x}}\subset\vect{x}$. The reader can visualise a lattice  in the $x$-$y$ plane of Figure \ref{fig1} with nodes $\vect{x}$  covering all red dots $\obs{\vect{x}}$. 

We are now in a position to stage the play and apply Bayes' theorem for a posterior distribution over all unknowns. First we  make a simple augmentation of the likelihood: $p(\obs{\vect{c}}|\vect{f},\mu_f,\nsigma)\equiv p(\obs{\vect{c}}|\vect{f}+\mu_f,\nsigma)$ with $\mu_f$ allowing for a constant (non-zero)   mean of the Gaussian process. The joint posterior over hyperparameters and  function values is then
\begin{equation}\label{eqPost}
p(\vect{f},\kappa,\mu_f,\nsigma|\mathcal{D}) = \frac{1}{Z}\underbrace{ p(\hat{\vect{c}}|\vect{f},\mu_f,\nsigma) }_\text{likelihood} \underbrace{ p(\vect{f}|\kappa) }_{\vect{f}-\text{prior}} \underbrace{ p_h(\kappa,\mu_f,\nsigma) }_\text{hyperprior}
\end{equation}
where $Z$ is a normalising constant. The prior over $\vect{f}$ is induced by the Gaussian process, $p(\vect{f}|\kappa) = \mathcal{N}(\vect{f}|\vect{0},\K_{\kappa})$ with a $N\times N$ covariance matrix with elements $[\K_{\kappa}]_{i,j}=k_{\kappa}(x_i,x_j)$ for all input pairs in $\vect{x}$. 
For the hyperprior over $(\mu_f,\nsigma)$ we also assume independent sigmoid Gaussians.

Calibration in our probabilistic model amounts to posterior inference about $\vect{f}$. Since the posterior distribution  is analytically intractable, we perform approximate inference by generating samples from \eqref{eqPost} (Section \ref{secMCMC}). From such a sample, we can extract the mean, mode  or {maximum a posteriori} (MAP) surface for stand-in replacements of ``classical'' deterministic point-estimates. What more is, we obtain a fully probabilistic representation  of local volatility. This is beneficial for several reasons: 
(i) posterior credible intervals provide a probabilistic notion of uncertainty for our estimates, (ii) the posterior uncertainty may be consistently incorporated in derivative  quantities of the model, 
$g(\sigma)$, through Bayes' estimator $\int g(\vect{f})p(\vect{f}|\mathcal{D})d\vect{f}$ and, (iii) it provides consistent probabilistic prediction of local volatility at unseen inputs $\pred{x}\in\mathcal{X}$ by
\begin{equation}\label{eqPred}
p(f(\pred{x})|\mathcal{D}) = \int \underbrace{ p(f(\pred{x})|\vect{f},\kappa) }_\text{cond. prior} \underbrace{ \frac{p(\obs{\vect{c}}|f(\pred{x}),\vect{f},\mu_f,\nsigma)}{p(\obs{\vect{c}}|\vect{f},\mu_f,\nsigma)} }_\text{likelihood ratio} \underbrace{p(\vect{f},\kappa,\mu_f,\nsigma|\mathcal{D})}_\text{posterior} d (\vect{f},\kappa,\mu_f,\nsigma).
\end{equation}
The conditional prior in \eqref{eqPred} follows from the predictive equations of Gaussian processes, i.e., $\mathcal{N}(f(\pred{x})|\vect{k}_{\kappa}^{\top}\K_{\kappa}^{-1}\vect{f},k_{\kappa}(\pred{x},\pred{x})-\vect{k}_{\kappa}^{\top}\K_{\kappa}^{-1}\vect{k}_{\kappa})$ with $\vect{k}_{\kappa} = \{k_{\kappa}(\pred{x},x)\}_{x\in\vect{x}}$  a column vector  \cite{rasmussen2006gaussian}. The likelihood ratio does not cancel out since \eqref{eqll} is non-factorisable over the local volatility surface: a model price $C( \{\vect{f},\pred{f}\},\obs{x})$  depends on all elements of $\{\vect{f},\pred{f}\}$ even if the dependency is strongest on values taken {locally} around $\obs{x}$. To first order, we may approximate it to be constant, in particular when we predict at a point $\pred{x}$ lying outside of $\obs{\vect{x}}$. It is then possible to generate predictions 
with a Gaussian mixture approximating  \eqref{eqPred},  or by direct  simulation.


\subsection{Inference algorithm}\label{secMCMC}
Models with latent variables like \eqref{eqPost}  are   intractable in general. 
We use Markov chain Monte Carlo to represent the posterior with samples. These are asymptotically exact methods in that they provide samples from a distribution which is arbitrarily close to the target distribution in the long run \cite{robert2004monte}. 

We generate states of $(\vect{f},\kappa,\mu_f,\nsigma)$ by Gibbs sampling of three blocks. First, we update $\vect{f}$ by targeting its conditional posterior 
\begin{equation*}
p(\vect{f}|\kappa,\mu_f,\nsigma,\mathcal{D}) \propto { p(\hat{\vect{c}}|\vect{f},\mu_f,\nsigma) } p(\vect{f}|\kappa).
\end{equation*}
We use \textit{elliptical slice sampling} for this purpose \cite{murray2010elliptical}. Second, we update $\kappa$ with  \textit{surrogate data slice sampling} \cite{murray2010slice}. 
This means targeting the conditional
\begin{equation}\label{eqSDSS}
p(\kappa|\boldsymbol{\eta},\vect{g},\mu_f,\nsigma,\mathcal{D}) \propto p(\hat{\vect{c}}|\vect{f}(\kappa,\boldsymbol{\eta},\vect{g}),\mu_f,\nsigma)p(\vect{g}|\kappa)p_h(\kappa)
\end{equation}
where $\vect{g}$ augments the model with a noisy version of $\vect{f}$, and $\vect{f}(\kappa,\boldsymbol{\eta},\vect{g})$  is a representation of $\vect{f}|\vect{g}$ with a standard normal variable $\boldsymbol{\eta}\sim\mathcal{N}(\vect{0},\mathbb{I})$---see \cite{murray2010slice} for details. Since \eqref{eqSSG} defines a transformed Gaussian, $\kappa_i = \kappa_{i,\text{max}}(1+\exp(-\vect{z}_i))^{-1}$ with $\vect{z}\sim\mathcal{N}(\vect{0},\mathbb{I})$, we target \eqref{eqSDSS} with elliptical slice sampling  in $\vect{z}$-space. Third, we target the conditional of $(\mu_f,\nsigma)$
\begin{equation*}
p(\mu_f,\nsigma|\kappa,\vect{f},\mathcal{D}) \propto p(\hat{\vect{c}}|\vect{f},\mu_f,\nsigma)p_h(\mu_f,\nsigma)
\end{equation*}
and again we use elliptical slice sampling for updating the corresponding bivariate Gaussian.

In terms of computational complexity, the second update of covariance parameters is the most expensive with a cost $\mathcal{O}(N^3)$ for (standard) inversion of the covariance matrix each time a new $\kappa$ is proposed. Proposing a new $\vect{f}$ also costs $\mathcal{O}(N^3)$ for the Cholesky decomposition of $\K_{\kappa}$, although the same decomposition can be recycled for consecutive proposals. The third update is the cheapest since it only requires computing the likelihood. One such evaluation involves solving $I$ systems of $J\times J$ tridiagonal matrices for the Crank-Nicolson scheme of the call price, where $I$ ($J$) is the number of maturities (strikes) of the input set $\vect{x}$  i.e., a cost  $\mathcal{O}(IJ)=\mathcal{O}(N)$. In fact, we can readily exploit that $\vect{x}$ is a Cartesian product and use Kronecker methods  when computing matrix decompositions and inversions \cite{saatcci2012scalable}. This reduces the cost to $\mathcal{O}(N^{3/2})$ for both the second (for each $\kappa$ proposal) and the first update (recycled throughout a $\vect{f}$-update with elliptical slice sampling).

\section{Related literature}
To the best of our knowledge, there is no previous work that treats  local volatility calibration 
with a full Bayesian approach based on functional priors. In fact, we  try to achieve what Hamida and Cont \cite{cont2004recovering} deemed too complicated just over a decade ago
``The computational complexity of the Bayesian approach has prevented it from being applied beyond a simple case such as Black-Scholes.''
 While a Bayesian analysis of Black-Scholes model is given in \cite{jacquier2000bayesian}, the authors propose an evolutionary algorithm to generate a population of local volatility. 
 All surfaces which, on average, reproduces benchmark prices within an average 
 bid-ask spread are treated as equal calibrates, and used to span  price-intervals of derivatives.  They use cubic splines  to represent the surfaces, a parametric model of  local volatility surfaces adopted before by \cite{jackson1998computation}. 
 
 The spline  model is also used by \cite{gupta2014robust} in a Bayesian approach to estimate its parameters. Prior beliefs  are here encoded by penalising local volatility deviations from the earliest at-the-money implied volatility with respect to  the Sobolev norm of first order. Hence, only first order differentiability is considered as a smoothness requirement. By defining the negative log-prior  proportional to this distance they obtain a Gaussian over the parameters of the spline model. They take the least-squares error to define a Gaussian likelihood with  (noise) standard deviation fixed at the average bid-ask spread. Similar to \cite{cont2004recovering} they  assign zero likelihood to surfaces which do not respect the bid-ask spread criterion. 

\begin{figure}
\centering
\includegraphics[scale=0.5,trim=  0 65 25 50,clip]{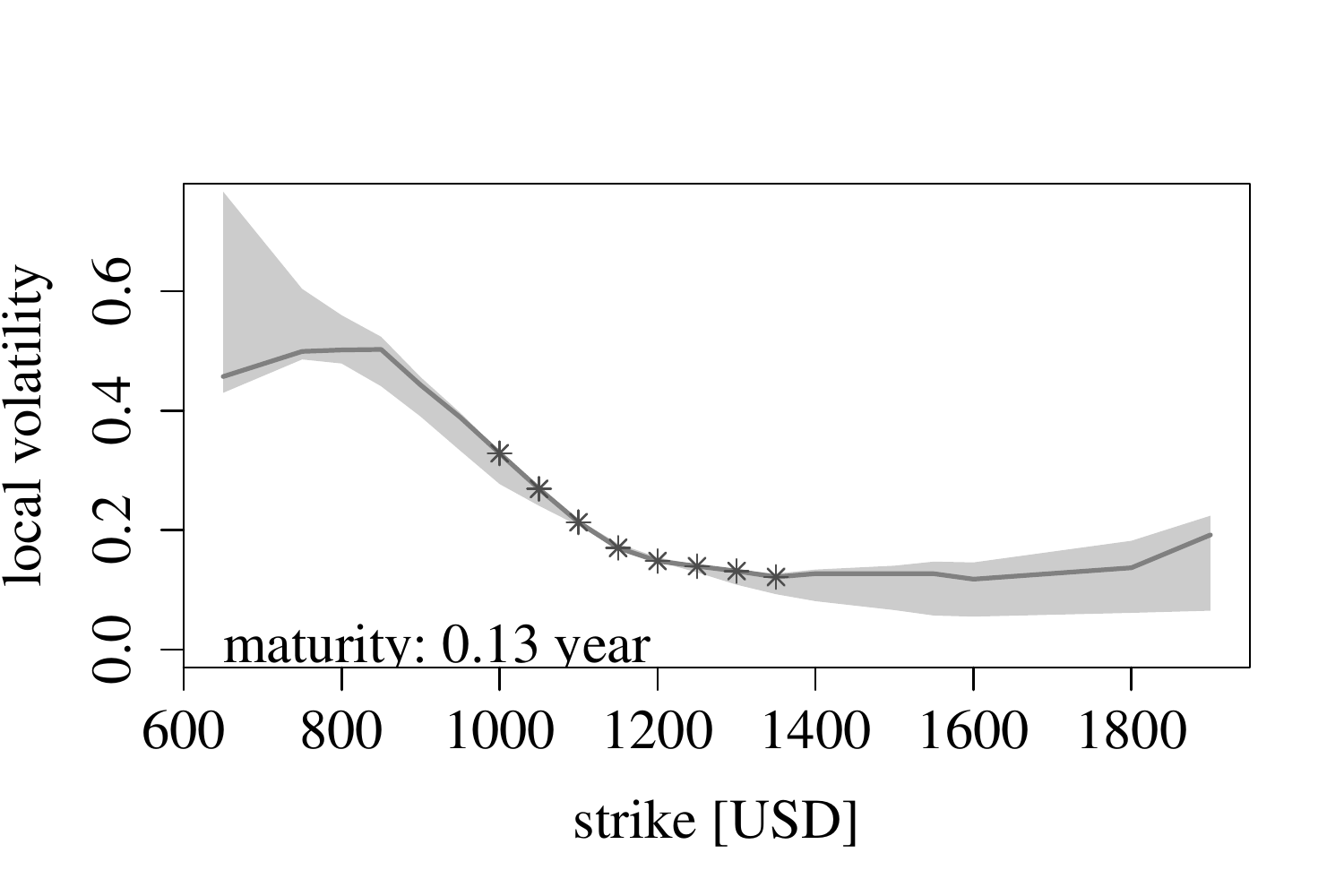} 
\includegraphics[scale=0.5,trim=46 65 25 50,clip]{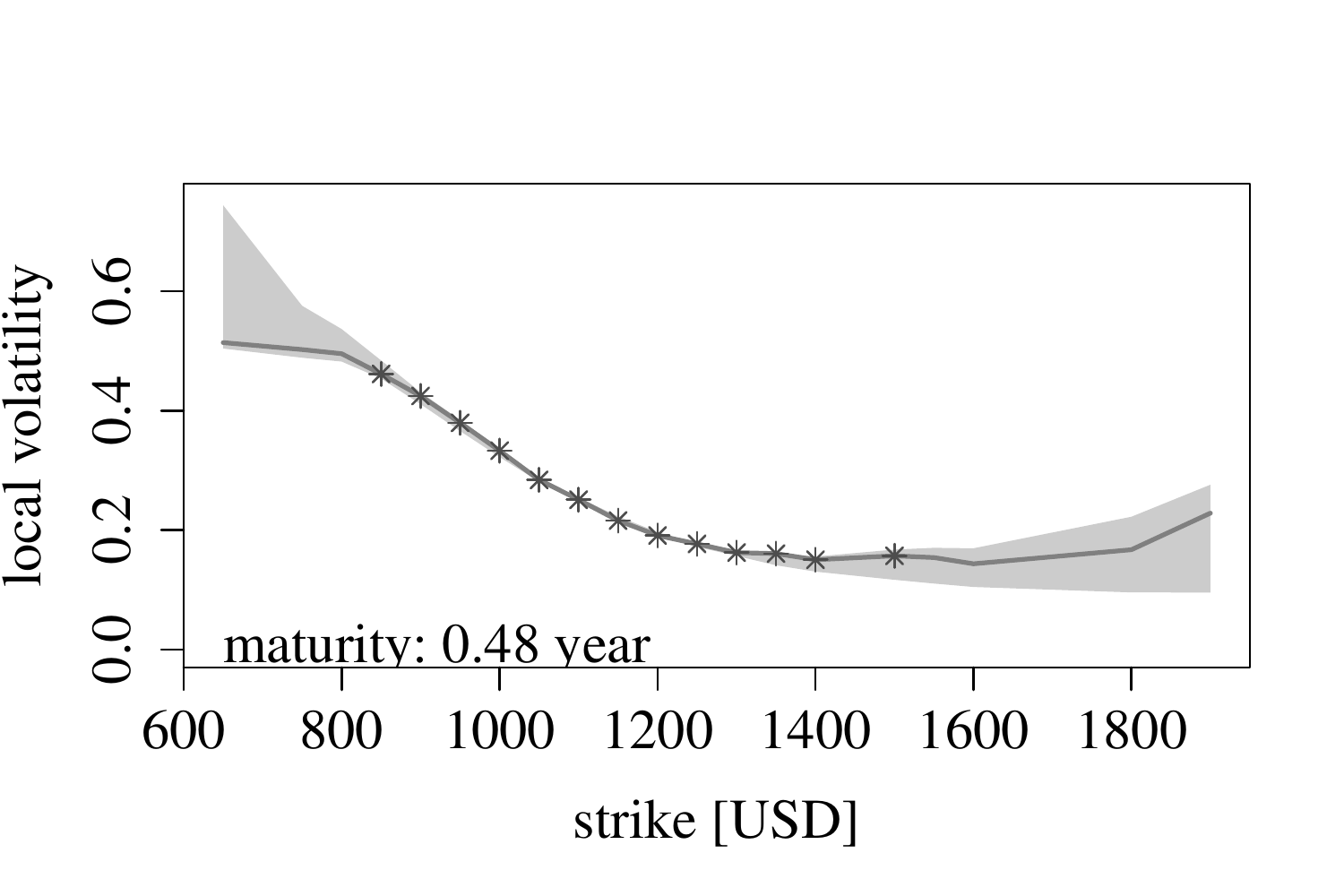} \\
\includegraphics[scale=0.5,trim=  0 10 25 50,clip]{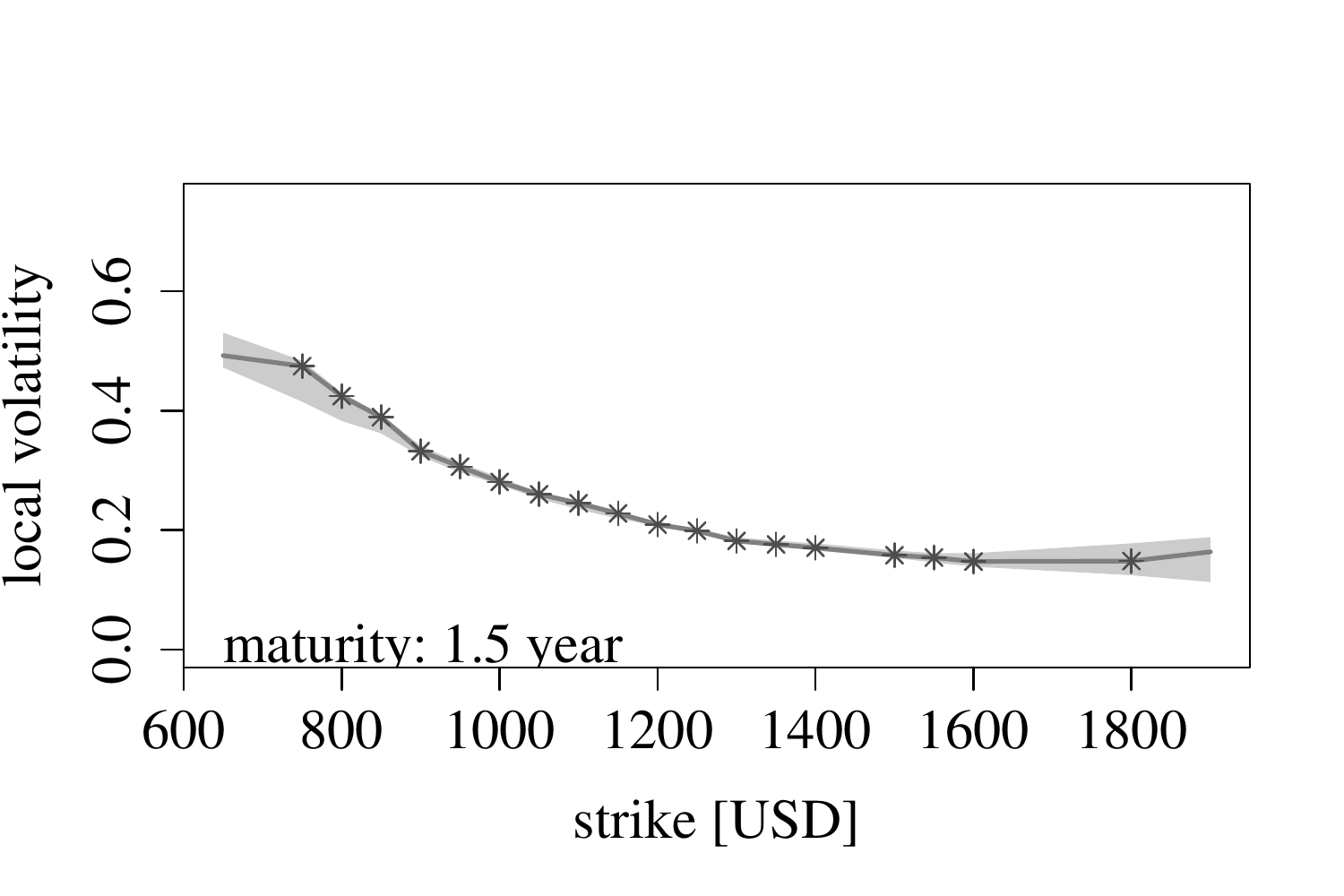} 
\includegraphics[scale=0.5,trim=46 10 25 50,clip]{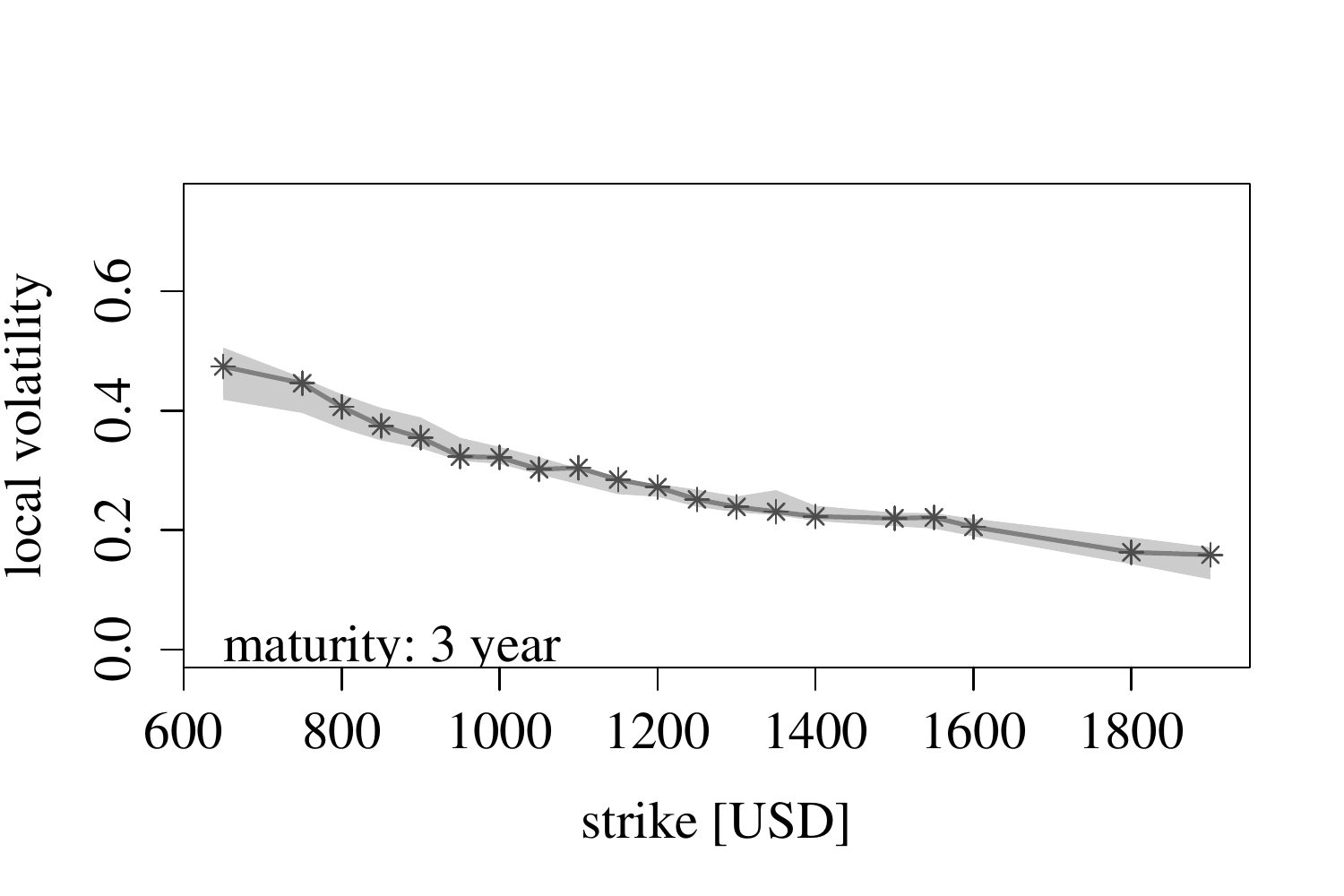} 
\caption{Posterior distribution over  local volatility as represented by the  maximum a posteriori surface (solid grey line) and a  credible interval of $\pm2$ standard deviations around the sample mean. 
 Asterisks show ``pseudo-volatilities''  to indicate where market values would be located.   The panels exhibit cross-sections taken at four maturities: 0.13, 0.48, 1.5 and 3 years.}
\label{fig2}
\end{figure}

On a conceptual level, our approach is similar to \cite{gupta2014robust}. The major difference is that we use a nonparametric \textit{functional} prior and perform \textit{full} Bayesian analysis.  While the Gaussian process defines a random function with properties specified by the covariance kernel, cubic splines are fundamentally deterministic interpolation polynomials, even if a distribution over the nodes induces a distribution over functions. Interpretability of the  models is therefore different:  encoding prior knowledge in a Gaussian process is straightforward by choosing an appropriate covariance (this may even be part of the learning process, see e.g. \cite{wilson2015kernel}). In the spline model, the choice of norm and attached Gaussian is  ad hoc such that prior beliefs about the polynomial function are somewhat indirect.  Moreover, the interpretation of hyperparameters is clear in the Gaussian process model. We include these in the inference process---and stress their importance---for a full Bayesian analysis. In contrast, pre-specified constants control the prior behaviour of the spline model, with an ex post robustness analysis of their impact on  model outputs---not on the posterior over local volatility.  Similarly, the impact of the fixed likelihood variance is analysed in terms of output robustness with respect to the data-noise (not  the variance parameter itself). Here we emphasise that noise variance 
do indeed affect the posterior over local volatility to a great extent---so do all hyperparameters. If we fix a corresponding $\nsigma$ 
from  bid-ask spreads and exclude it  from inference, we obtain much wider confidence regions compared to  Figure \ref{fig2}.

Finally we just mention that there is a large body of literature on calibration through regularised problems of the form \eqref{eqG}, e.g., \cite{chiarella2000calibration, crepey2003calibration,egger2005tikhonov, jackson1998computation}. Albeit deterministic, these are  related to our approach:  unconstrained minimisation of \eqref{eqG} corresponds  to maximising \eqref{eqll} (for fixed $\nsigma$), i.e., the maximum likelihood estimator---or a completely uninformative prior. Similarly, minimising  \eqref{eqG} with a quadratic regularisation term corresponds to maximising \eqref{eqPost} (for fixed hyperparameters), i.e., the MAP estimator.


\section{Experiments}
We demonstrate our approach  with market  data from call options on the S\&P 500 index, shown in Figure \ref{fig1}. The left figure shows the (best) bid and ask price of  473  options quoted on 4 January 2010. The mid-market prices are converted to implied Black-Scholes volatilities, shown in the right figure. {These two representations are equivalent as the Black-Scholes price is monotonically increasing in implied volatility.} The latter expresses option prices in a standardised  ``currency'',  often preferred by option traders. It can be though of as  representing a  \textit{market sentiment} in that it is the market's expectation of future fluctuations in the underlying stock-price.  A high level of implied volatility attached to a strike-maturity yields an expensive option. It anticipates a volatile  stock-price (high risk), such that trading in options to protect from (adverse) price movements will be expensive.     

\paragraph{Calibration}
 From the market data, we take a subset $\mathcal{D}$ of $n=117$ options for calibration with strike-maturities $\hat{\vect{x}}$ distributed over 20 strikes and 8 maturities. These are marked as blue strike-maturity points in  right Figure \ref{fig1}. The input set $\vect{x}$ thus forms a  $20\times8$ grid covering these points.  This is  to have a reasonably sized input of $N=160$ points for computational efficiency in the MCMC algorithm (compared with $103\times14=1442$ for the full  data). We still cover the original ranges of strikes and maturities; we just take a less dense input set evenly distributed over the original quotes.

To set-up an operational model, we scale $\vect{x}$ to lie in the unit square and set $\kappa_{i,\max}=1$, $\forall i$, for the covariance hyperprior, and $\sigma_{\epsilon,\max}=0.5$ and  $ \mu_{f,\max} = \log 0.5$ for the likelihood . These values stem from  heuristics and some initial runs of the MCMC algorithm. Length-scales around one (maximum) give  slowly varying volatility surfaces much smoother than the implied volatility in right Figure \ref{fig1}: prior $\kappa$-ranges $(0,1)$  provide enough flexibility. For the likelihood, the maximum noise standard-deviation  0.5 is in the same range as the average bid-ask spread of 2.5. Note, however, that the noise models the deviance of the mid-market price from the (unobserved) fair price, not the (half) bid-ask spread. 

We run 50,000 iterations of the three-step algorithm described in Section \ref{secMCMC}. We discard the first 10,000 states as burn-in, and subsample the reminder to obtain a thinned posterior sample of 1000 local volatility surfaces. The result is represented in Figure \ref{fig2}, where  we  show  four cross-sections of the volatility surfaces taken along the strike dimension, for maturities of 0.13, 0.48, 1.5 and 3 years. 
In each panel, we show a slice from the MAP surface and a credible interval of $\pm2$  standard deviations around the mean (estimated from the posterior sample point-wise at each $x$-input). 

The shapes of the credible intervals are symptomatic of two main factors:   non-linearity of the pricing/likelihood function and data availability. For early maturities (0.13 and 0.48 years), the price is not very sensitive to model parameters when looking at 
strikes far from  the current asset price  $S_t = \$1130$. For $K>>S_t$, there is not much chance of the option reaching at/in-the-money 
before maturity, i.e., $K\leq S_T$. This keeps the price close to zero (since the option is likely to expire worthless; see left Figure \ref{fig1}). Similarly, for $K<<S_t$ there is little chance of going out-of-money and the option's price is close to its intrinsic value, $S_t-K$ (its time value is small). In effect, the price insensitivity makes the likelihood  relatively uninformative about local volatility over these points. 
 For the  same reasons there is  little trading in the options. Market quotes are concentrated around at-the money, $K=S_t$, 
which also contribute to increased uncertainty outside this region as there is lack of data. The effects can be seen as a high degree of uncertainty for large and small strikes in the top panes of Figure \ref{fig2}.

On the other hand,  posterior confidence is high at strikes close to at-the-money where the price is most sensitive to  local volatility, and where there is a high concentration of quotes. For late maturities (3 years) the price is also sensitive to model parameters  for small and large strikes. With time, there is an increasing probability that the option will leave its in- or out-of-money region. Considering the dynamics \eqref{eqLVSDE}, local volatility is a parameter of diffuseness of the stock-price. For long maturities, the volatility level therefore has higher impact on option prices  over all strikes. In effect, the credible region of bottom right Figure \ref{fig2} shows a rather constant uncertainty across strikes. Compared to earlier maturites, the band is  wider around at-the-money. Another reason for this is  data availability with a larger  gap between maturities of 2 and 3 years (see Figure \ref{fig1}).

The  posterior sample of local volatility surfaces gives a probabilistic picture of the local volatility  function. This is interesting in its own right as it provides information on what can be learned (and not learned) from data about the latent function, i.e., to what degree of certainty one can pin down $\sigma$-values over different strike-maturity points. The result---point-estimate(s) and associated uncertainty---can then be used when employing the model for other purposes. For example, if pricing a derivative which is sensitive to $\sigma$ over points with high calibration uncertainty, one better takes this uncertainty into account for a robust price-estimate. 


 \begin{figure}
\centering
\includegraphics[scale=0.5,trim=  0 65 25 50,clip]{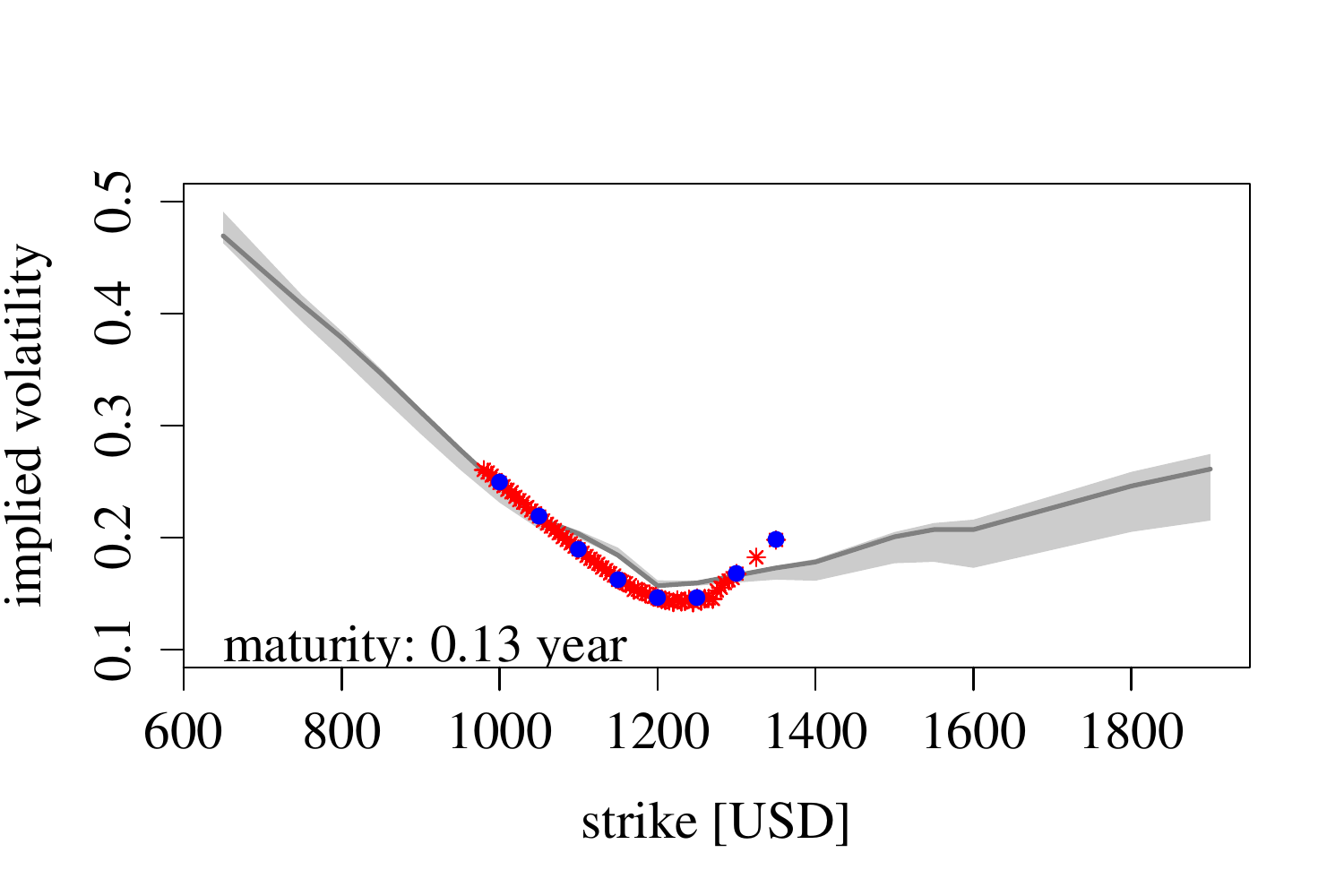} 
\includegraphics[scale=0.5,trim=46 65 25 50,clip]{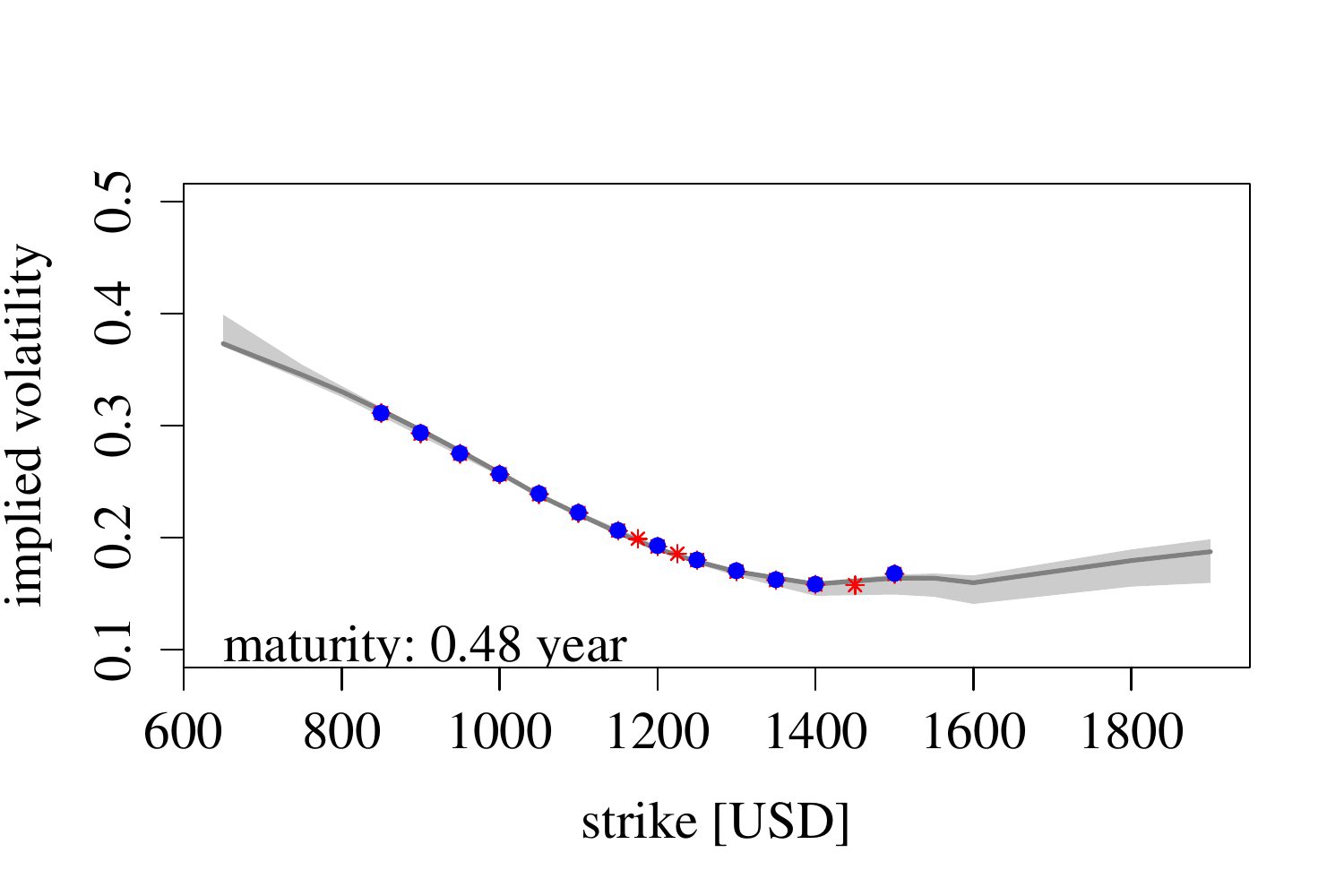} \\
\includegraphics[scale=0.5,trim=  0 10 25 50,clip]{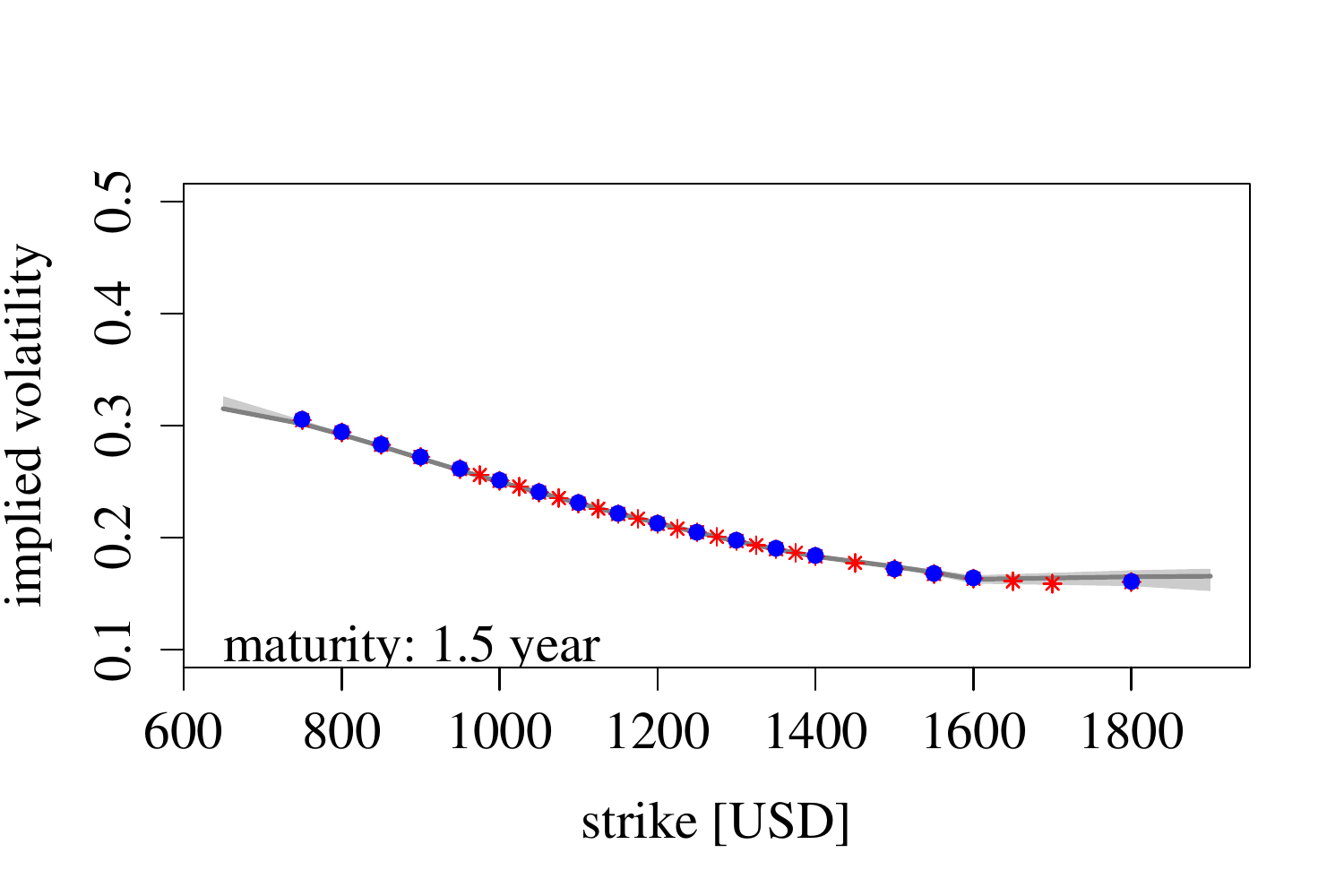} 
\includegraphics[scale=0.5,trim=46 10 25 50,clip]{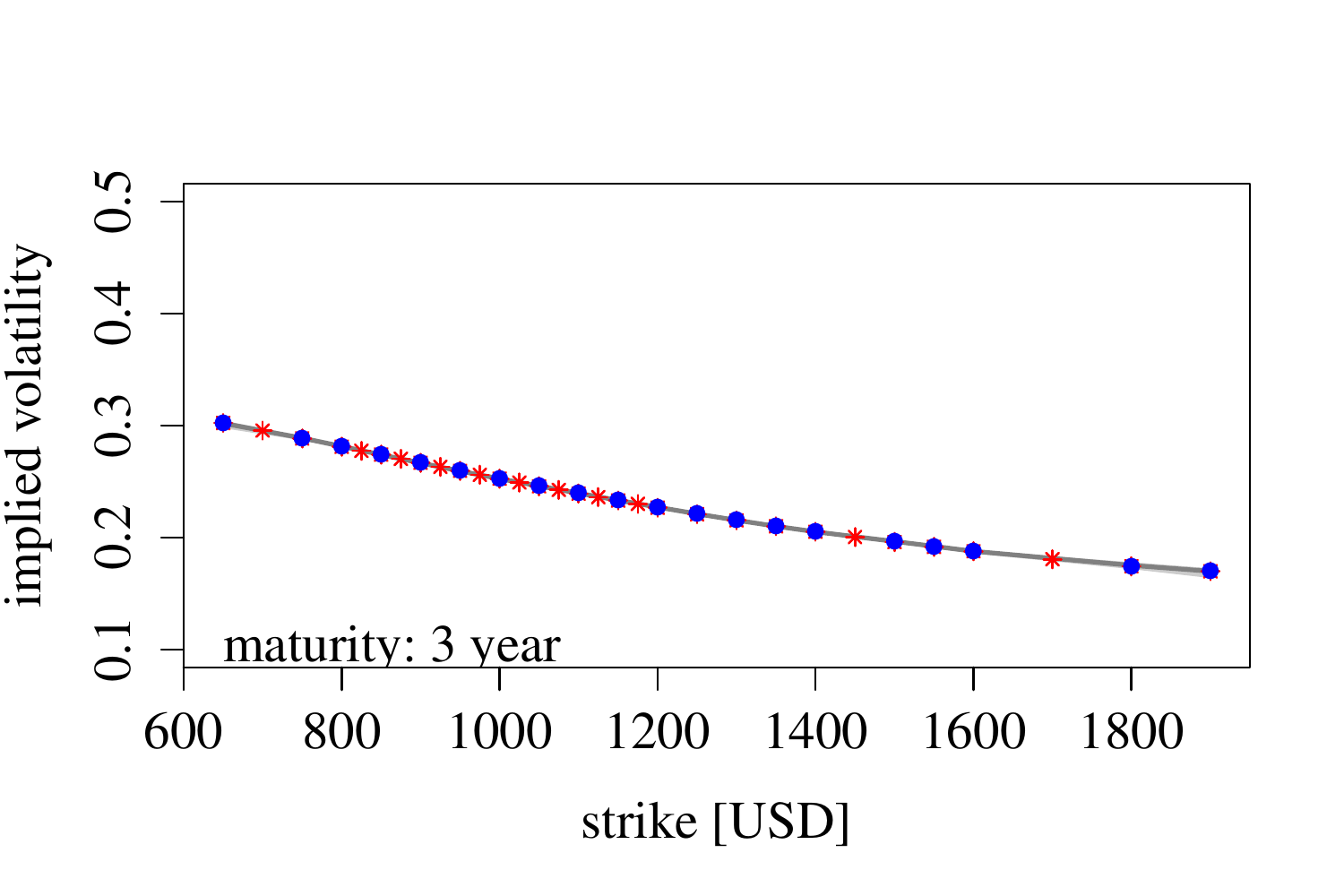} 
\caption{Posterior distribution over  implied volatilities: maximum a posteriori surface (solid grey line) and  $\pm2$ standard deviations around the  mean (light grey area). Estimates calculated from the transformed  sample of local volatility surfaces (Figure \ref{fig2}). Calibration data $\mathcal{D}$  plotted with blue dots  while red asterisks show the excluded market data. }
\label{fig3}
\end{figure}

\paragraph{Re-pricing market data}
 To see how well the calibrated model explains market data we transform the posterior sample over local volatility to samples over call prices and implied volatilities. The result is shown in Figure \ref{fig3} for implied volatilities; corresponding call prices (not shown for brevity) show confidence regions which are completely tight (prices are \$0--500 with standard deviations $\sim$ \$0.25), with all market prices indistinguishable from the MAP surface. 
 In terms of implied volatilities, Figure \ref{fig3} show that the MAP surface is  close to the data    used for calibration (blue dots) and to the market data excluded from this set (red asterisks). The only exemptions are some implied volatilities at the shortest maturity, as shown in the top left pane of Figure \ref{fig3}. This confirms the fact that  the local volatility model is often inconsistent with market prices  at short maturities \cite{gatheral2011volatility}. An explanation is that the underlying stock price is modelled as a continuous process, while the market anticipate jump-like behaviour which contribute to more pronounced implied-volatility skews (as observed in top-left Figure \ref{fig3}). 
 
 Implied volatility MAP-to-market errors are within $0.0003\pm0.02$ and within  $0\pm0.01$ if excluding the shortest maturity; both indicate a strong re-pricing performance of our approach (recall that the likelihood is based on prices, not implied volatilities) and competitive to what is reported in the literature: \cite{cont2004recovering} report  $\sim0\pm0.04$ for a set of DAX options, \cite{luo2010local}  a few percentage points. 
 
 Finally we note that the uncertainty of the  local volatility estimate  transfers to  modest credible intervals for implied volatility, especially over the in-the-money region. The reason is, again,   model price/implied volatility  (in)sensitivity to the local volatility parameter---even if there is a high level of local-volatility uncertainty over some strike-maturity regions, it is not propagated back to uncertainty over  prices/implied volatilities. This property of call options should, however, not be taken for granted for other derivatives, especially if they are designed to be heavily dependent on volatility.
 
\begin{figure}
\makebox[\textwidth][c]{  
\includegraphics[scale=0.36,trim=60 50 70 50,clip]{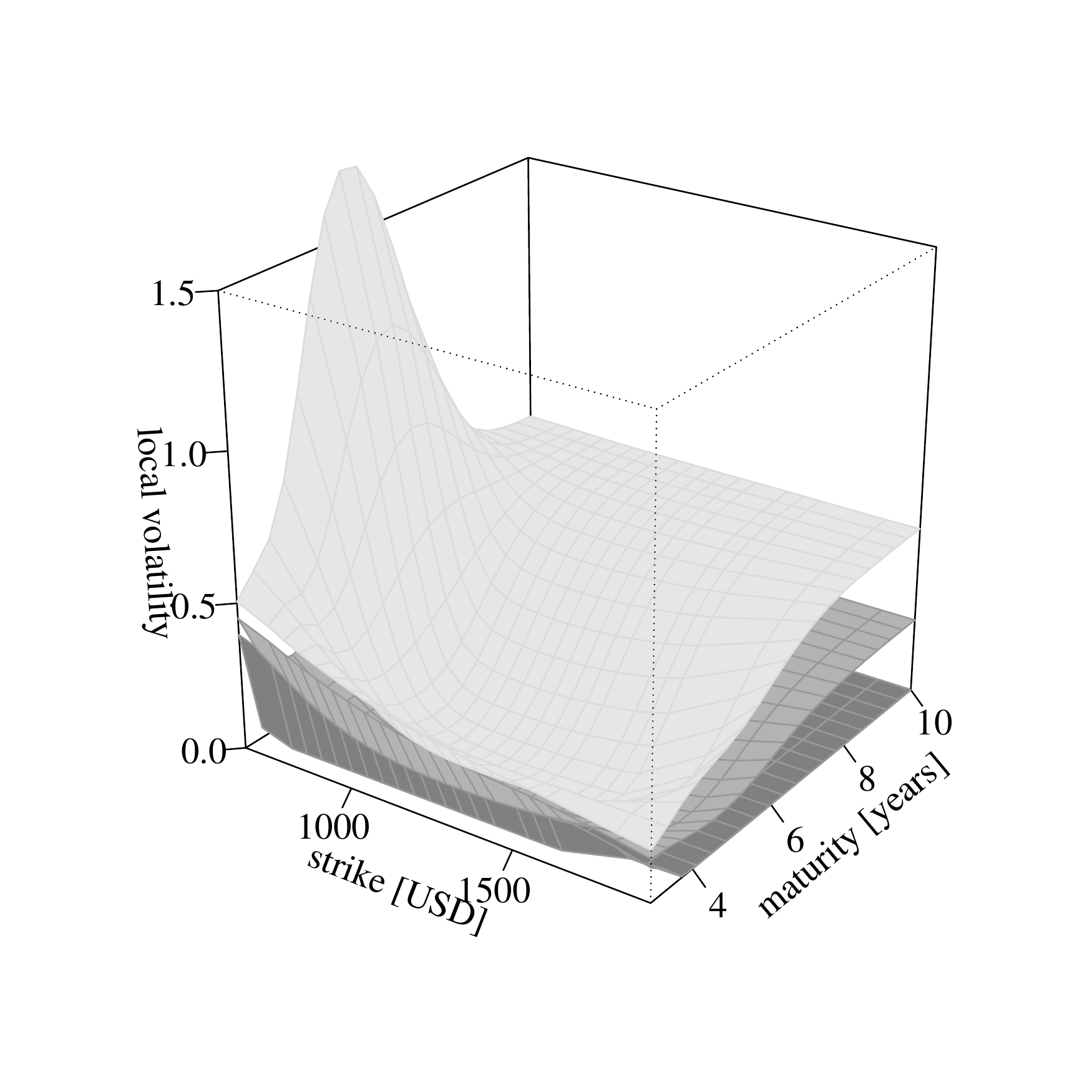}
\includegraphics[scale=0.36,trim=60 50 70 50,clip]{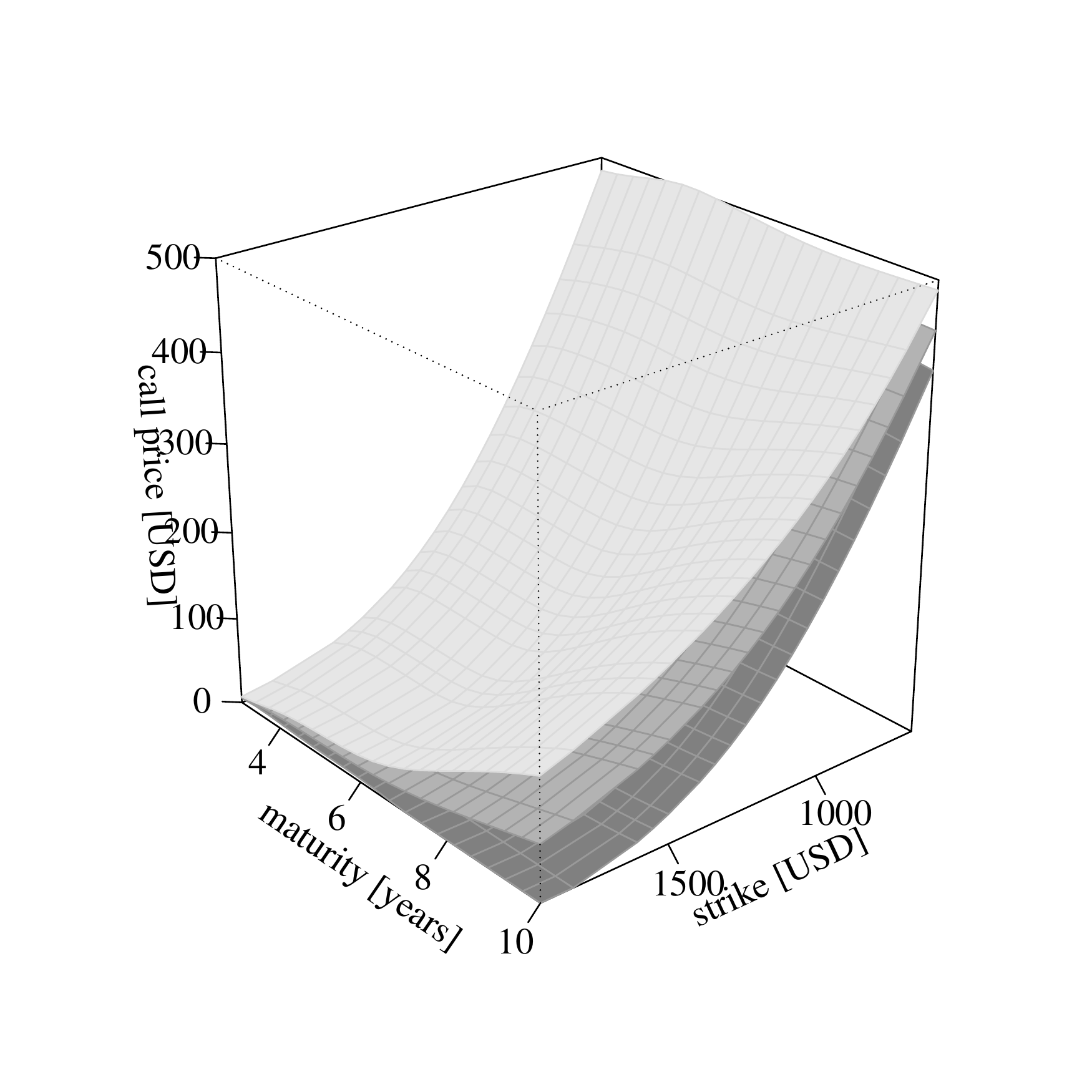}
\includegraphics[scale=0.36,trim=60 50 70 50,clip]{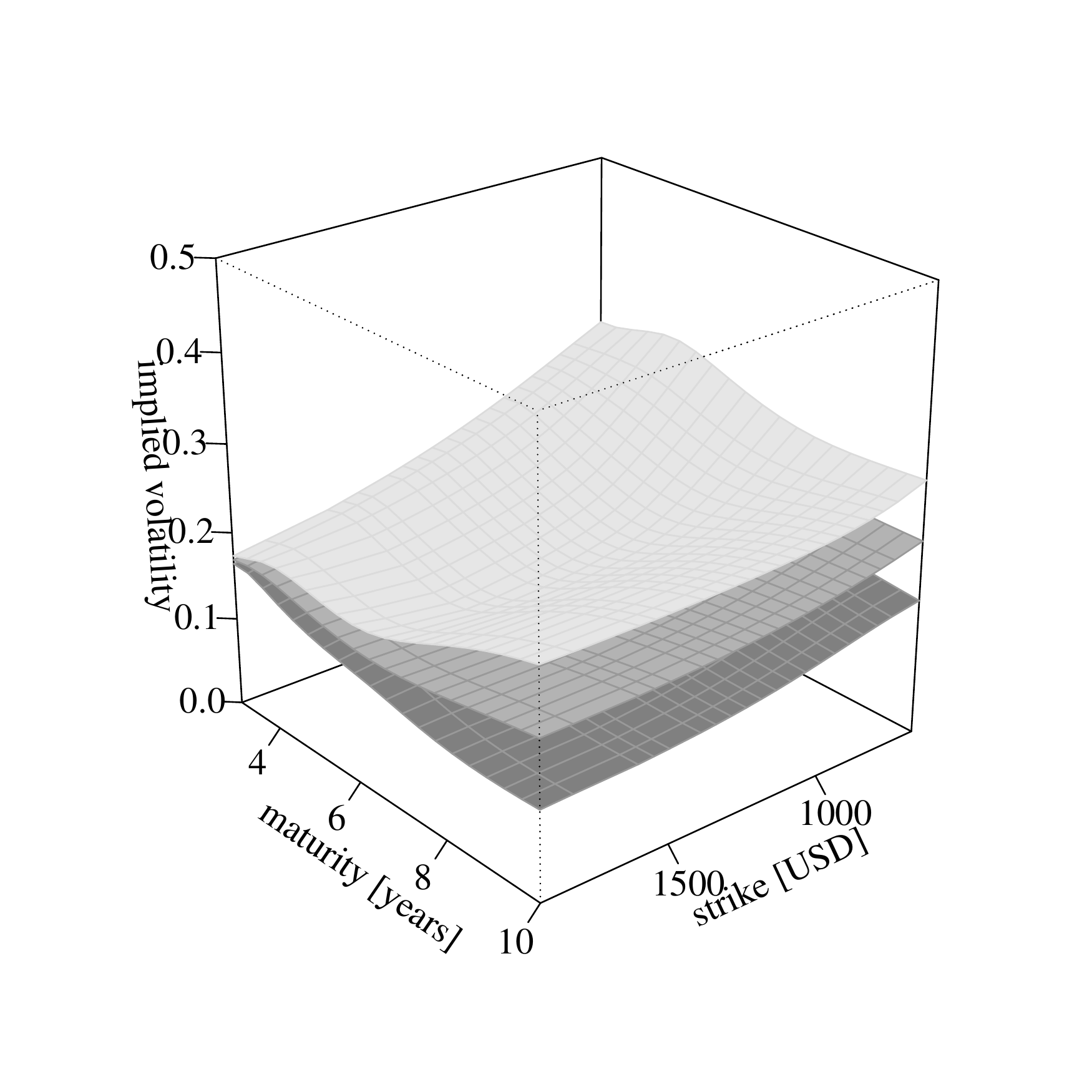}
}
\caption{\textbf{Left}  Posterior predictive distribution over local volatility for  maturities of 4--10 years: credible interval of $\pm2$ standard deviations around the mean.  \textbf{Middle } Predicted call prices. The figure is rotated to better show prediction uncertainty for  long-dated options. \textbf{Right } Implied volatilities corresponding to predicted call prices. Compared to the calibration (Figure \ref{fig3}) the predictions uncertainty is generally higher   (same $z$-axis scale in both figures) and notably increasing in maturity.  } 
\label{fig4}
\end{figure}

 
 \paragraph{Prediction} As  mentioned, it is common to employ a calibrated model to price off-exchange  products, i.e., possibly non-standard derivatives not (publicly) traded. These may dependent on unquoted strike-maturity points which require  interpolation and extrapolation of the calibrated local volatility surface.  
 We readily have a machinery for this purpose that provides a  predictive posterior over local volatility at any strike-maturity point of our choice. As an illustrative example we consider long dated options with maturities 4--10 years. We generate a prediction sample of 1000 local volatility surfaces by simulation: each of 100 randomly selected states $(\vect{f},\kappa,\mu_f)$ from the posterior sample are used to calculate the moments of the conditional prior in \eqref{eqPred}. We then generate  10 surfaces 
 from every such set of moments. The result is shown in left Figure \ref{fig4} as a point-wise credible envelope of $\pm2$ standard deviation around the mean. Comparing with the calibration  (Figure \ref{fig2}) the prediction uncertainty is substantial over small/large strikes and increasing in maturity up to a point $\sim7$ years, from which the surfaces level off to their prior confidence $\sim 0.2\pm0.3$.      The sample is used to a predict call prices (middle Figure \ref{fig4}) which in turn are transformed to a sample over implied volatilities (right Figure \ref{fig4}). Both illustrates how prediction uncertainty of local volatility propagate: long-dated options  can only be consistently priced subject to an increasing  uncertainty-range which is  substantial for the latest maturities.

\section{Conclusion}

We have presented an approach for Bayesian inference in the local volatility model based on  Gaussian process priors. To the best of our knowledge, this is the first full Bayesian analysis with a nonparametric  functional prior. We discussed several benefits of our approach such as straightforward encoding of prior beliefs and interpretability,  the gain of a probabilistic notion of calibration uncertainty, posterior predictions consistent with data, and the evaluation of derivate model-quantities with  posterior information taken into account. We also proposed an MCMC algorithm for inference. For real market data, we demonstrated encouraging calibration performance and discussed how the pricing model and benchmark data interact to generate the resulting posterior, with focus on uncertainty. Finally, we predicted local volatility over unseen, long maturities and used the result to generate a predictive distribution for long-dated call options.

As immediate extensions of this work, one might consider alternations of the setup and model. A prior Matérn covariance would probably achieve higher likelihood values (smaller re-pricing errors) to the cost of less smooth volatility surfaces. In terms of the likelihood, the noise model \eqref{eqll} is the simplest choice. Market features such as liquidity and transaction costs could motivate more intricate models with structural and also non-Gaussian noise. Such  could readily be adopted in the likelihood and their parameters added to the inference process. We also note the freedom to choose other calibration benchmarks---bid and ask prices for example---as long they permit a pricing operator which can be evaluated (Monte Carlo  is a go-to method, although involving SDE simulation in the the likelihood  would be terrible in terms of computational effort). Several extensions exist  for the local volatility model itself---such as stochastic interest rates and discrete dividends--- which could be investigated. A more fundamental and interesting extension is to consider a  Bayesian Gaussian-process  approach to the \textit{Brunick-Shreve model} \cite{brunick2013mimicking}. The latent diffusion function of this model is extended to depend on a third state, the running maximum of $S$.  Since the payoff of barrier options depends on the running maximum, barriers take a similar role for this model as calls have for  local volatility. With a pricing equation corresponding to \eqref{eqDupire} derived in \cite{hambly2016forward}, we could thus mimic our prevailing approach with an extended input space of the Gaussian process.  We leave this  for future work.

\bibliography{nips_2018}
\bibliographystyle{abbrv}

\end{document}